%% file: main.tex
\documentclass[letterpaper,twocolumn,10pt]{article}
\usepackage{usenix,epsfig,endnotes}
\usepackage{array} 
\usepackage{multirow} 
\usepackage{amsmath}
\usepackage{amssymb}
\usepackage{hyperref}

\begin{document}

\date{}

\newcommand{\VulnScopper}{{\em VulnScopper }}


\title{Unveiling Hidden Links Between Unseen Security Entities }
 
\author{
    {\rm Daniel Alfasi}\\
    Reichman University, Israel
\and
    {\rm Tal Shapira}\\
    The Hebrew University of Jerusalem, Israel
\and
    {\rm Anat Bremler Barr}\\
    Tel Aviv University, Israel
}

\maketitle

\thispagestyle{empty}

\subsection*{Abstract}
The proliferation of software vulnerabilities poses a significant challenge for security databases and analysts tasked with their timely identification, classification, and remediation. With the National Vulnerability Database (NVD) reporting an ever-increasing number of vulnerabilities, the traditional manual analysis becomes untenably time-consuming and prone to errors. This paper introduces \VulnScopper, an innovative approach that utilizes multi-modal representation learning, combining Knowledge Graphs (KG) and Natural Language Processing (NLP), to automate and enhance the analysis of software vulnerabilities. Leveraging ULTRA, a knowledge graph foundation model, combined with a Large Language Model (LLM), \VulnScopper effectively handles unseen entities, overcoming the limitations of previous KG approaches.

We evaluate \VulnScopper on two major security datasets, the NVD and the Red Hat CVE database. Our method significantly improves the link prediction accuracy between Common Vulnerabilities and Exposures (CVEs), Common Weakness Enumeration (CWEs), and Common Platform Enumerations (CPEs). Our results show that \VulnScopper outperforms existing methods, achieving up to 78\% Hits@10 accuracy in linking CVEs to CPEs and CWEs and presenting an 11.7\% improvement over large language models in predicting CWE labels based on the Red Hat database.
Based on the NVD, only 6.37\% of the linked CPEs are being published during the first 30 days; many of them are related to critical and high-risk vulnerabilities which, according to multiple compliance frameworks (such as CISA and PCI), should be remediated within 15-30 days. We provide an analysis of several CVEs published during 2023, showcasing the ability of our model to uncover new products previously unlinked to vulnerabilities. As such, our approach dramatically reduces the vulnerability remediation time and improves the vulnerability management process.

\section{Introduction}
Effective vulnerability management involves continuously identifying, classifying, and remediation of vulnerabilities in software systems. To effectively manage vulnerabilities, security databases provide valuable information about vulnerabilities, weaknesses, and products and ensure they maintain accurate and up-to-date information. Security databases, such as the National Vulnerability Database (NVD)\cite{NVD}, gather information about vulnerabilities in different software products and libraries. Meanwhile, companies' CVE databases like the Red Hat CVE Database\cite{RedHatCVEDatabase} specifically collect records of Red Hat products and open-source software vulnerabilities. In 2023, the National Vulnerability Database (NVD) reported 28,830 vulnerabilities publicly, and this number has been increasing annually. 

Security analysts manually analyze this vast amount of vulnerability records, classify them into weaknesses, and identify vulnerable products. This time-consuming, error-prone analysis is crucial for security teams to address vulnerabilities. The categorization of Common Vulnerabilities and Exposures (CVE) to Common Weakness Enumeration (CWE) provides a blueprint for security teams to understand the core flaws and weaknesses of a software product. Identification of Common Platform Enumerations (CPEs) within CVEs allows security experts to understand the scope of a vulnerability by specifying the vulnerable software products. In reality, it takes days, weeks, and even months to discover all vulnerable products from the time a CVE is first recorded. Software vulnerabilities such as Log4Shell\cite{log4shell, CVE-2021-44228} have been updated with a new CPE configuration over 247 days (when this paper was written), leaving software products that used these vulnerable libraries exposed for an extended period. Our analysis finds that more than 6\% of the CVEs recorded between 2019 and 2023 in NVD received a new CPE configuration update more than 30 days after being recorded. 

The difficulty in promptly updating CVE records with accurate CPE information arises from the complex task of determining which products incorporate the vulnerable component. Given the sheer volume of software products and their dependencies, tracing the vulnerable library's usage is often not straightforward. It is essential to have a tool that can accurately identify relevant CPEs to speed up the update process. By leveraging machine learning, such a tool could significantly improve analysts' efficiency. Such a tool would allow them to focus on critical tasks instead of manually searching through records to pinpoint affected products.

To address this issue, the research community has recently applied natural language processing (NLP) to automate the analysis of software vulnerabilities. V2W-BERT\cite{V2W-BERT} used vulnerability and weakness descriptions to determine the type of weakness, \cite{cpe_Labeling_of_CVE_Summaries} employed descriptions from vulnerabilities to determine their product, \cite{KeyAspectsVulnerabilityReports} extracted key aspects from vulnerability descriptions using regular expression patterns. As pointed out by \cite{CWE-CVE-CPE-Relations}, a limitation of NLP-based approaches is that each CVE entry is treated separately, making implicit relations hard to identify. To capture the rich relation information of security entities, \cite{CWE-CVE-CPE-Relations} suggested a knowledge graph (KG) reasoning approach. \cite{DeepWeek}, \cite{EmbeddingPredictingSWEntity} and \cite{HybridGat} employed a hybrid representation approach using a KG and entity description. Despite the promising results, traditional KG approaches employ shallow embedding lookup mapping for each entity, making them unsuitable for predicting unseen entities. 

We show that none of the current methods meet all the requirements for identifying the scope of vulnerabilities. These requirements include accurately determining the weaknesses and software products affected by the vulnerability, the scale to handle the vast amount of vulnerability data that can boot model performance, and the ability to handle new vulnerabilities without additional training.

In this paper, we present \VulnScopper, a multi-modal representation learning approach that uses KG and NLP to meet specified requirements. \VulnScopper is built on the top of the vulnerability knowledge graph, allowing it to learn the rich relation information between security entities. To address the limitation of handling unseen entities in previous KG approaches, we utilize ULTRA\cite{galkin2023ultra}, a graph foundation model that can generalize to graphs with unseen nodes. To enhance link prediction accuracy, \VulnScopper's multi-modal approach uses embeddings of security entity descriptions obtained from OpenAI Ada, an NLP embedding model.

We implement and evaluate \VulnScopper on two security datasets, NVD and Red Hat CVE database, in two modes, transductive and inductive. The transductive mode evaluation assumes that all entities are seen during training. In contrast, the inductive mode assumes the presence of new vulnerabilities during the evaluation that were not seen during the training process. According to our evaluation results, \VulnScopper achieved a 70\% and 71\% ranking accuracy on Hits@10\footnote{Hits@10 measures the percentage of relevant items that appear within the top 10 results.} for linking CVEs and CWEs on RedHat and NVD datasets, respectively. Additionally, when ranking links between CVEs and CPEs, \VulnScopper achieved 76\% and 71\% Hits@10 accuracy on RedHat and NVD datasets, respectively. Furthermore, we demonstrate \VulnScopper's ability to successfully adapt the different CWE labeling of the RedHat dataset compared to Large Language Models such as ChatGPT4\cite{openai2023chatgpt4}, resulting in an 11.7\% improvement in Hits@10 accuracy for predicting CWE.

Our contributions are summarized below:
\begin{enumerate}
    \itemsep 0pt
    \item We present a novel approach to connect between CVEs to CWEs and CPEs. \VulnScopper combines graph and text representation learning approaches to predict the links between these entities accurately.
    \item We solve the scalability issues of the previous graph-based methods using ULTRA, allowing \VulnScopper to deal with the sheer amount of data.
    \item We demonstrate \VulnScopper's ability to connect unseen vulnerabilities, a previous limitation of knowledge graph-based methods. This allows security teams to focus on vulnerabilities without having to retrain models.
    \item We evaluated \VulnScopper on two different datasets in transductive and inductive modes, and the results demonstrate its ability to achieve high prediction accuracy, outperforming previous work.
    \item We demonstrated \VulnScopper's power by discovering new links between recent vulnerability cases such as 'libwebp,' 'libcurl', and 'Log4shell' to products. We manually analyze the results for each case to validate the prediction accuracy.
\end{enumerate}

\section{Preliminaries}
\input{sections/background}

\section{\VulnScopper Motivation}
\input{sections/problem_statement}

\section{Related Work}
\input{sections/related-work}

\section{ \VulnScopper Design}
\input{sections/vulnscopper}

\section{Experimental Results} 
\input{sections/experiments}

\input{sections/evaluation}

\input{sections/results} 

\section{Case studies: Link Prediction of Unseen Entities}
\input{sections/examples}

\section{Discussion}
\input{sections/discussion}

\section{Conclusions}
\input{sections/conclusions}

{\bibliographystyle{acm}
\bibliography{main}}

\section{Appendices}
\appendix 
\input{sections/appendix}

\end{document}

%% file: sections/background.tex
In this section, we provide relevant background information on security entities, CVEs, CWEs, and CPEs. We then delve into the background of Natural Language Processing (NLP) and Knowledge Graph (KG) techniques. Specifically, we discuss the Knowledge Graph Link prediction problem, which is essential to the core of our work. Finally, we introduce ULTRA, a knowledge graph reasoning model used by \VulnScopper, and discuss its relevant details to our algorithm.

\subsection{Security Entities}
\label{sec:security_entities}
The Common Vulnerabilities and Exposures (CVE) are at the core of vulnerability management, which provides a standardized reference for publicly known information-security vulnerabilities and exposures. The CVE system facilitates data sharing across various security tools and services, ensuring a universal understanding of specific vulnerabilities. 

\subsubsection{ Common Weakness Enumeration (CWE) }
\label{sub-sec: cwe}
Complementing the CVE identifiers is the Common Weakness Enumeration (CWE) framework\cite{CWE_MITRE}, which categorizes software and hardware weakness types. CWE aids in presenting a standardized language for describing software security weaknesses in architecture, design, or code, which adversaries exploit. Understanding the CWE context of a vulnerability provides insights into the potential root causes and, by extension, guides the development of more secure software practices and tools. 

\begin{figure}[h!]
\centering
\includegraphics[width=0.5\textwidth]{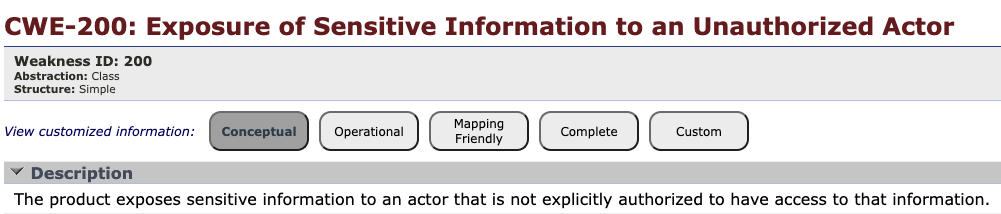}
\includegraphics[width=0.5\textwidth]{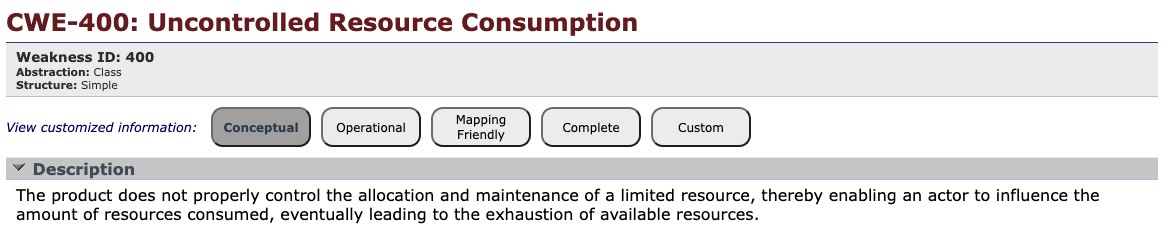}
\caption{CWE-200 and cwe-400 from cwe.mitre.org.}
\label{fig:cwe_desc_example}
\end{figure}

For example, the CWE-200 description in figure \ref{fig:cwe_desc_example}, "The product exposes sensitive information to an actor that is not explicitly authorized to have access to that information" and CWE-400 description also in figure \ref{fig:cwe_desc_example}, "The product does not properly control the allocation and maintenance of a limited resource, thereby enabling an actor to influence the amount of resources consumed, eventually leading to the exhaustion of available resources."

The CWE database exhibits a hierarchical structure where weakness types are organized in a parent-child relationship, with broader categories serving as parents to more specialized child weaknesses. At the apex of this hierarchy, parent weaknesses provide general vulnerability descriptions, while children represent narrower and more defined types. For example, "CWE-74: Improper Neutralization of Special Elements ('Injection')" serves as a parent to "CWE-79: Cross-Site Scripting (XSS)", illustrating how general concepts are refined into specific vulnerabilities.

Additionally, the CWE database encompasses relational terms like "can precede" and "can follow," which articulate temporal or logical relationships between weaknesses. These terms indicate potential sequences in which specific weaknesses may lead to others, suggesting a progression or evolution of vulnerabilities that could be exploited in a chain. For instance, "CWE-20: Improper Input Validation" can precede "CWE-22: Improper Limitation of a Pathname to a Restricted Directory ('Path Traversal')", indicating that improper input validation may lead to a Path Traversal.

\subsubsection{ Common Platform Enumeration (CPE) }
The Common Platform Enumeration (CPE) is a structured naming scheme for information technology systems, software, and packages. Within vulnerability management, CPE names specify a particular vendor, product name, version, and other attributes associated with software components. For example the CPE record cpe:2.3:a:google:chrome:*:*:*:*:*:*:*:* indicates application type of product Chrome by Google and cpe:2.3:o:fedoraproject:fedora:37:*:*:*:*:*:*:* is operating system Fedora version 37.

By standardizing vulnerability categories and scope, security professionals can accurately determine the relevance of vulnerabilities to specific software systems. Using standardized systems can help security teams manage threats more efficiently and accurately, addressing them before attackers exploit them.

\subsection{Natural Language Processing}
NLP methods can extract and interpret complex patterns within textual information, enabling machines to understand, infer, and respond to human language meaningfully. Language models have evolved significantly in their application to link prediction, classification, and Named Entity Recognition (NER) tasks. 

Early rule-based methods, such as regular expressions, were adept at capturing explicit patterns in text but often needed to catch up due to their rigidity and the complexity of language \cite{regex_usage}. The introduction of statistical methods such as TF-IDF \cite{TF-IDF-text-classification} has enabled the use of more refined text features, which are utilized for tasks like text classification. Word2vec \cite{word2vec} embeddings represent a breakthrough in capturing semantic similarities and enabling a wide range of downstream tasks. With their deep contextual understanding, transformer-based \cite{attention_is_all_you_need} approaches have set new benchmarks across various NLP tasks, outperforming previous methods. 

Large Language Models (LLMs) \cite{llm_survey} like OpenAI's ChatGPT, Ada embedding model, and Facebook's LLAMA have revolutionized NLP tasks by providing robust, scalable solutions that can understand and generate human-like text. These LLMs, trained on diverse datasets, can perform complex tasks such as link prediction, text classification, and NER with remarkable accuracy, often surpassing human performance benchmarks. VulnScopper leverages OpenAI's Ada LLM to derive detailed embeddings from descriptions of security entities, which capture the nuanced semantics beyond traditional feature extraction methods.

\subsection{Knowledge Graphs Reasoning}
Knowledge graph reasoning methods enable the extraction of insights by interpreting the links between entities, facilitating complex reasoning over a network of interconnected data. Knowledge graph reasoning has evolved from the foundational PageRank \cite{pagerank}, which ranks nodes in a graph based on their link structure, to sophisticated node representation techniques like node2vec \cite{node2vec} that capture a node's neighborhood information. 

Knowledge graph embedding methods such as TransE \cite{bordes_transe}, DistMult \cite{distmult}, and RotatE \cite{rotate} advanced the field by embedding entities and relations into continuous vector spaces, facilitating tasks like link prediction. The introduction of Graph Attention Networks (GAT)\cite{GAT} incorporated the attention mechanism, allowing for more refined entity-relation representations. 

Nodepiece \cite{galkin2022nodepiece} has emerged as an effective approach for inductive link prediction. It leverages the relational context of a node to generate its representation. Recent models such as NBFNet \cite{nbfnet} and ULTRA \cite{galkin2023ultra} can generalize graphs for unseen entities and relations. Additionally, ULTRA acts as a foundation model that allows for link prediction based on a pre-trained ULTRA instance.

\subsubsection{Knowledge Graph Link Prediction}
\input{sections/problem_formulation}

\subsubsection{ULTRA}
\input{sections/ultra}
\label{sec:ultra}

%% file: sections/problem_formulation.tex
Link prediction is an essential task in graph reasoning, enabling the prediction of relations between entities. In the context of security entities, this capability allows us to automate the association of CVEs with CWEs and CPEs, thereby enhancing the understanding of vulnerabilities and their corresponding implications across different platforms and weaknesses.

Formally, let a Knowledge Graph be denoted as $G$, characterized by a set of entities $E$, a set of relations $R$, and a set of triplets $T$. Each triplet $t = (e_1, r, e_2)$ represents a relationship $r \in R$ between two entities $e_1, e_2 \in E$, where $e_1$ and $e_2$ are the head and tail entities, respectively, and $r$ is the relation connecting them. The objective of KG link prediction is to infer the missing head or tail entity in a given triplet, either predicting the tail entity in the triplet $t = (e_1, r, ?)$ or the head entity in the triplet $t = (?, r, e_2)$. To facilitate effective link prediction, $G$ is divided into graphs $G_{train}$, $G_{val}$, and $G_{test}$, represented by their respective triplets.

In the \emph{transductive setup}, it is assumed that all entities $E$ and relations $R$ are known during the training phase on $G_{train}$. During the inference phase on $G_{test}$, the model tries to predict previously unseen triplets.

Conversely, the \emph{inductive setup} does not presuppose complete knowledge of all entities during the training phase. This approach introduces an inference graph, denoted as $G_{inference}$, which extends the dataset with links that were not seen during training. In this setup, the training and inference graphs differ, symbolized as $G_{train} \neq G_{inference}$. This distinction requires the model to generalize beyond the entities encountered in $G_{train}$, to effectively handle entirely new entities present in $G_{inference}$.

%% file: sections/ultra.tex
To allow VulnScopper to learn the KG entity representation while dealing with unseen vulnerabilities, we use ULTRA\cite{galkin2023ultra}. ULTRA is a step towards foundation models for KG, aiming to learn universal and transferable graph representations. It achieves this by conditioning relational representations on their interactions, enabling pre-trained models to generalize and fine-tune across unseen KGs with any relation vocabularies. ULTRA combines three main components: 1) Relation graph construction, 2) Conditional Relation Representation, and 3) Entity-level link prediction.

\begin{figure}[h!]
\centering
\includegraphics[width=0.5\textwidth]{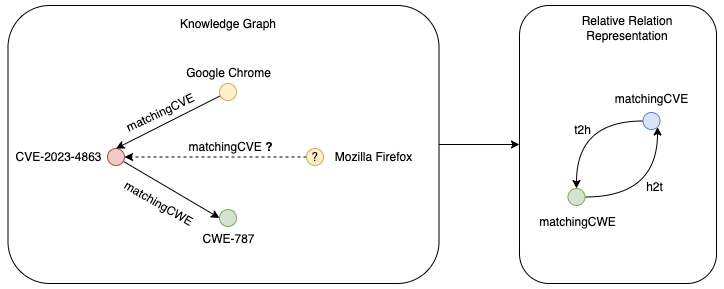}
\caption{ULTRA's Relation graph construction step on our vulnerability KG.}
\label{fig:ultra_rel_graph}
\end{figure}

In the Relation graph construction step, Graph $G=(E, R, T)$ is transformed using a lifting function $G_r=LIFT(G)$ to create a new graph $G_r$, where relations become nodes. This graph distinguishes between four fundamental types of relational interactions illustrated in Figure \ref{fig:ultra_rel_graph}: tail-to-head (t2h) edges, head-to-head (h2h) edges, head-to-tail (h2t) edges, and tail-to-tail (t2t) edges.

Subsequently, the conditional relation representation component creates a $d$-dimensional node representation $R_q\in\mathbb{R}^{|\mathcal{R}| \times d}$ of $G_r$ conditioned on query $q=(h,r,?)$. The conditioning is implemented through $INDICATOR_r$ function and employs a Graph Neural Network (GNN) over 
$G_r$:
\[
h_{v|u}^0 = INDICATOR_{r}(u, v, q) = 1_{u=v} * 1^d
\]
\[
h_{v|u}^{t+1} = UPDATE(h_{v|u}^t, AGG(MSG(h_{w|u}^{t}, R_{fund})))
\]
The indicator function assigns a vector filled with ones to the node $u$ if it matches the query relation $q$, and assigns a zero vector to all other nodes. The GNN architecture is based on NBFNet \cite{nbfnet} with a DistMult\cite{distmult} message ($MSG$) function and sum aggregation. Each unique relation $r\in \mathbb{R}$ in the query has its own matrix of conditional relation representation $R_q\in \mathbb{R}^{|R| \times d}$ that entity-level link predictor can use.

In the final phrase, an additional NBFNet operates on the entity level and was utilized to predict links at the entity level. The process starts by initializing the head node with the query vector while other nodes are set to zero. Each GNN layer then applies a non-linear function $g^t(\cdot)$ to refine relation representations used in the MESSAGE function $MSG$:
\[
h_{v|u}^0 = INDICATOR_{e}(u, v, q) = 1_{u=v} * R_q[q]
\]
\[
h_{v|u}^{t+1} = UPDATE\left( h_{v|u}^t, AGG( MSG(h_{w|u}^t, g^{t+1}(r))) | r\in R \right)
\]
The function $g^t(\cdot)$ uses a two-layer MLP with ReLU activation. Sum aggregation and a linear layer update these representations, culminating in an MLP that converts node states to a probability score, indicating the likelihood of a node being the tail of the query.

%% file: sections/problem_statement.tex
In this section, we will analyze the association between CVEs, CPEs, and CWEs and demonstrate the challenges of manually analyzing these associations, which is the motivation to 
 \VulnScopper. 
 
\subsection{CVE and CPE Association}

When a CVE is published, the complete list of relevant CPEs that are vulnerable is not immediately known, and it sometimes takes months to reveal the true impact. The list of CPEs is updated over time to reflect this.

We have analyzed more than 111,000 CVEs, which are all the CVEs that were published from 2019 to 2023 by NVD. Among these years, the average number of CPEs a CVE has is 3. Additionally, we observed that the number of recorded CVEs that received a new CPE configuration after one day, seven days, 30 days, and 180 days is 6.83\%, 6.69\%, 6.09\%, and 4.83\%, respectively. Table \ref{new_cpe_vul} shows the percentage of vulnerabilities updated with a new CPE after 1,7,30 and 180 days of creation per year. We note that the percentage of new CPE configurations has been lower for CVEs in recent years as their scope has yet to be fully revealed. 

\begin{table}[h!]
\centering
\begin{tabular}{cccccc}
\hline
Year & Total & 1 Day & 7 Days & 30 Days & 180 Days \\
\hline
2019 & 18,352 & 9.10\% & 8.99\% & 8.53\% & 6.93\% \\
2020 & 18,522 & 9.73\% & 9.70\% & 9.54\% & 8.26\% \\
2021 & 19,771 & 9.50\% & 9.34\% & 8.15\% & 4.29\% \\
2022 & 25,260 & 6.79\% & 6.72\% & 5.86\% & 1.73\% \\
2023 & 29,549 & 1.84\% & 1.70\% & 1.43\% & 0.28\% \\
\hline
\end{tabular}
\caption{ Vulnerabilities updated with a new CPE after 1, 7, 30 and 180 days from their creation}
\label{new_cpe_vul}
\end{table}

\subsection{CVE and CWE Association}
\label{sub-sec:cve_cwe_association_analysis}
Missing CWE information can delay security experts' understanding of a vulnerability's underlying weakness and lead to misguided prioritization or patching strategies, ultimately compromising the overall security posture.
Our analysis finds that among all CVEs available in the NVD database, 25.39\% have missing CWEs, while in the Red Hat CVE database, 42.33\% are missing. 

This gap is not merely a matter of missing data but also arises from the inherent categorization approach of each database. With its broader scope, NVD looks at all vulnerable products, while vendor-specific databases like Red Hat focus solely on their products, leading to discrepancies in the CWEs assigned to the same CVEs. Among all CVEs presented in both databases and categorized with a CWE, 44.3\% have a different CWE. Furthermore, the hierarchical nature of CWEs allows for CVEs to be tagged with either general or specific weaknesses, potentially leading to confusion and additional workload for security teams who strive to understand and address the most precise type of weakness. For example, CVE-2023-38545 is categorized as CWE-787 in NVD \cite{CVE-2023-38545_nvd} while Red Hat uses  CWE-119 \cite{CVE-2023-38545_redhat}.

\subsection{The Exposure Time of Software Systems}
The exposure time of software systems refers to the period between the publication of a CVE record and the time security teams address it. This duration depends on two primary factors: First, the time taken for the scope of a CVE record to be fully understood, and Second, the time taken for the vulnerability to be resolved by the security teams. Since the second factor depends on the first factor and the security team's efficiency, it is crucial to address the first factor as quickly as possible to prevent software systems from being exploited.

Therefore, automating the process using a dedicated tool for linking CVEs to CWEs and CPEs is critical for the efficiency of security teams. Such a tool can rank related CWEs and CPEs for CVE entities and guide security analysts toward understanding the full scope of a vulnerability.

%% file: sections/related-work.tex
The utilization of vulnerability databases to gain insights into vulnerabilities and threats has been extensively studied in the hi-tech industry and academia.

\begin{figure}[h!]
\centering
\includegraphics[width=0.5\textwidth]{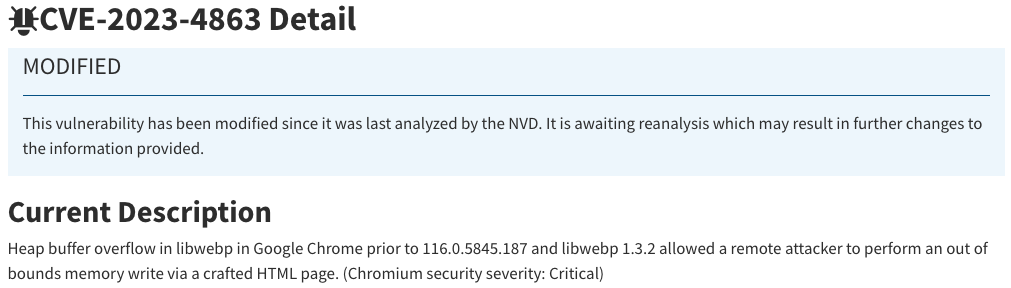}
\caption{CVE-2023-4863 vulnerability description, NVD.}
\label{fig:libwebp_desc}
\end{figure}

NLP-based methods use entity descriptions to gain an understanding of their connections, \cite{cpe_Labeling_of_CVE_Summaries} proposed an automated framework for matching CVE descriptions with their product through an NLP technique for the Named Entity Recognition (NER) task combined with a deep neural network to construct their matching CPE configurations.
In \cite{KeyAspectsVulnerabilityReports}, the authors used regular expressions to extract vulnerability aspects such as vulnerability type and products from CVE descriptions, then employed a Bi-LSTM model to predict the missing aspects from new vulnerabilities. V2W-BERT \cite{V2W-BERT} employs a transfer learning approach to automate the link prediction task between CVEs and CWEs. Although these works were relatively effective for CWE prediction, description-based approaches process each CVE individually. Therefore, they lose the vulnerability's relational information, such as associated CPEs, which are not part of its descriptions. Recent vulnerability cases, such as 'libwebp' CVE-2023-4863\cite{CVE-2023-4863}, affected a wide range of products; some of them, as shown in figure \ref{fig:libwebp_desc}, are not part of the vulnerability description such as Fedora, Debian Linux, Firefox, Microsoft Edge, and Mozilla Thunderbird. 

To deal with that issue, Graph and description-based approaches such as \cite{analysis_method_sec_kg} suggested creating a security vulnerabilities knowledge graph to improve the analysis and visualization of software vulnerabilities. In \cite{CWE-CVE-CPE-Relations}, the authors constructed a vulnerability knowledge graph using CVE, CWE, and CPE entities and employed a TransE\cite{bordes_transe} model to predict the links between them. The works of \cite{DeepWeek, EmbeddingPredictingSWEntity} utilized TransCat and TransH KG link prediction algorithms, respectively, where the concatenation of graph embeddings and description embeddings obtained from the Word2Vec\cite{word2vec} algorithm were employed. \cite{HybridGat} employed TransE for KG embedding and Word2Vec embeddings of the entity descriptions, concatenated both, and sent them into a graph attention network (GAT) \cite{GAT}. 

While these methods proved effective, they still inherit the limitation of being unable to predict links to unseen nodes. To the best of our knowledge, all previous vulnerability graph-based approaches employed models that use a lookup mapping for the embeddings of each entity and relation in the graph. In a real-life scenario where security teams need to address new vulnerabilities, these methods are impractical as it is only possible to predict links to new vulnerabilities by re-training the entire model.


\VulnScopper employs a multi-modal approach composed of a ULTRA for knowledge graph reasoning and OpenAI's Ada\cite{OpenAIAdaEmbeddings} embeddings for textual description.

%% file: sections/vulnscopper.tex
In this section, we present \VulnScopper, a multi-modal KG and NLP-based approach to predict links between CVEs to CWEs and CPEs. \VulnScopper leverages the structural and textual representations of the vulnerability knowledge graph entities for link prediction. We continue our discussion on constructing two vulnerability knowledge graphs based on NVD and Red Hat CVE databases and how they are utilized in \VulnScopper.

\subsection{Vulnerability Graph Construction}
\label{sub-sec:vul_graph_construction}
\begin{figure}[h!]
\centering
\includegraphics[width=0.5\textwidth]{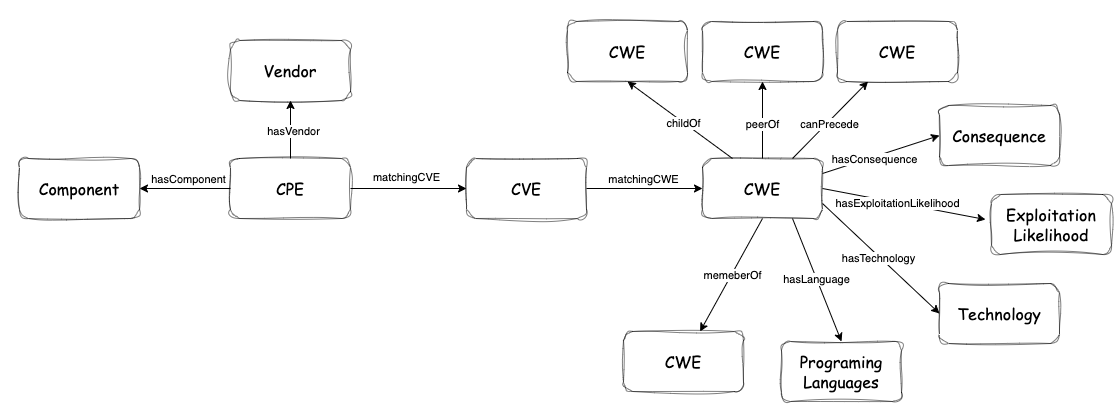}
\caption{The vulnerability knowledge graph schema used by \VulnScopper.}
\label{fig:graph_schema}
\end{figure}

To create the \VulnScopper vulnerability knowledge graph, we consider the different types of security entities defined in Section \ref{sec:security_entities} and their relations. The schema of our graph is shown in Figure \ref{fig:graph_schema}. The complete list of node types and relation types in \VulnScopper is shown in Table \ref{table:graph_node_rel_types}.

The first step in the graph construction is retrieving all the available vulnerability-related knowledge from the NVD API \cite{NVD_API} and the Red Hat CVE database API \cite{RedHat_SecurityDataAPI} up to October 2023. We parsed this data, initially presented in JSON format, to extract useful information about each CVE record. Each CVE record included its relevant CWE, vulnerable CPE configurations, and a vulnerability description. To retrieve the software vendor and component type of the CPE, we parsed it based on the CPE 2.3 structure [3]. To represent the product, we use the shortened CPE configuration. In example, \textit{cpe:2.3:o:debian:debian\_linux:12.0:*:*:*:*:*:*:*} is shorten to \textit{cpe:2.3:o:debian:debian\_linux} in order to represent Debian Linux product. We link each CPE configuration to the CVE record with the relation type "MatchingCVE", while each CPE connected to its vendor using the "hasVendor" relation and to its component using the "hasComponent" relation type.

Next, for each CVE that has CWE connected to it, we add a new relation called "MatchingCWE". 
We use MITRE CWE data \cite{mitre_cwe_download} to enrich the graph with additional contextual data with each CWE information. This data contains meaningful structural information about the rich hierarchy of CWEs as described in Section \ref{sub-sec: cwe}, and additional properties such as programming language, technology, exploitation likelihood, and consequence. We use the CWE hierarchical information to create four types of relations in our graph, including "childOf", "peerOf", "canPrecede", and "memberOf". Finally, we utilize the CWE properties obtained from MITRE CWE data to establish relations between CWEs and their matching properties "Language", "Technology", "Exploitation Likelihood", and "Consequence" as shown in Figure \ref{fig:graph_schema}.

\begin{figure}[h!]
\centering
\includegraphics[width=0.5\textwidth]{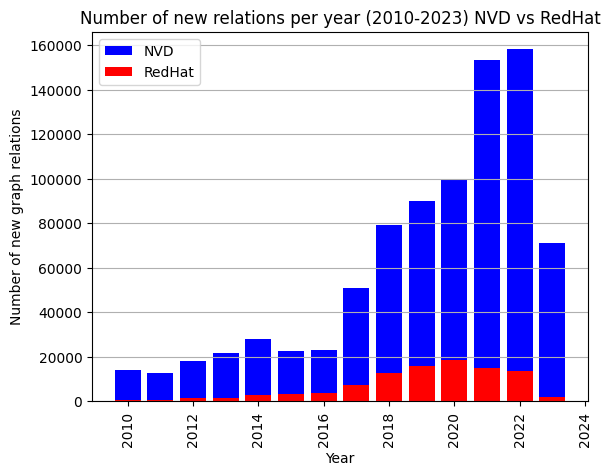}
\caption{Number of relations that are added to each graph per year.}
\label{fig:graph_links_per_year}
\end{figure}

The fast growth in the volume of data year by year is depicted in Figure \ref{fig:graph_links_per_year}, highlighting the complexity and size of the vulnerability knowledge graph constructed. To maintain the quality of our graph and increase prediction effectiveness, we apply a filter to exclude CVEs that lack associations with any CPE or CWE entries. Ultimately, the final graph used by VulnScopper employs 12 types of relations. Table \ref{table:graph_stats} shows the number of nodes and relationships.

\begin{table}[h!]
\centering
\begin{tabular}{|l|c|r|c|}
\hline
\textbf{Graph Version} & \textbf{\#CVEs} & \textbf{\#CPEs} & \textbf{\#CWEs} \\ \hline
NVD                   &         212,360       &       103,892      & 393   \\ \hline
RedHat                &         16,104       &       9,328      & 224   \\ \hline
\end{tabular}
\caption{The number of CVEs, CPEs, and CWEs in NVD and Red Hat graphs.}
\label{table:graph_entities_count}
\end{table}

\begin{table}[h!]
\begin{tabular}{|l|c|r|c|}
\hline
\textbf{Graph Version} & \textbf{\#Nodes} & \textbf{\#Rel Types} & \textbf{\#Total Rel} \\ \hline
NVD                   &         349,032       &       12      & 1,052,423   \\ \hline
RedHat                &         27,720       &         12    &  109,781  \\ \hline
\end{tabular}
\caption{The number of unique nodes, unique relation types, and relations of the NVD and Red Hat graphs.}
\label{table:graph_stats}
\end{table}

\begin{table}[h!]
\centering
\begin{tabular}{|c|l|}
\hline
\textbf{Element} & \multicolumn{1}{c|}{\textbf{Type}} \\ \hline
Node & CVE, CWE, CPE, Vendor, Component \\
 & Language, Technology, Consequence \\
 & Exploitation Likelihood \\ \hline
Relation & matchingCWE, matchingCVE, hasVendor \\
 & hasComponent, childOf, peerOf, \\
 & canPreceed, memberOf, hasLanguage, \\
 & hasTechnology, hasExploitationLikelihood, \\ 
 & hasConsequence \\ \hline
\end{tabular}
\caption{\VulnScopper's graph node and relation types.}
\label{table:graph_node_rel_types}
\end{table}

\subsection{Security Entities Description Utilization}
In \VulnScopper, we employ CVE descriptions sourced from the NVD and Red Hat databases and CWE descriptions from MITRE. For other security entities lacking explicit descriptions, their labels serve as substitutes; for instance, "cpe:2.3:a:google:chrome" is the description for the matching CPE entity. We then establish a mapping from each entity label to its corresponding description to facilitate its use during model training.

\subsection{Components of \VulnScopper}
\begin{figure*}[h]
    \centering
    \includegraphics[width=1\textwidth]{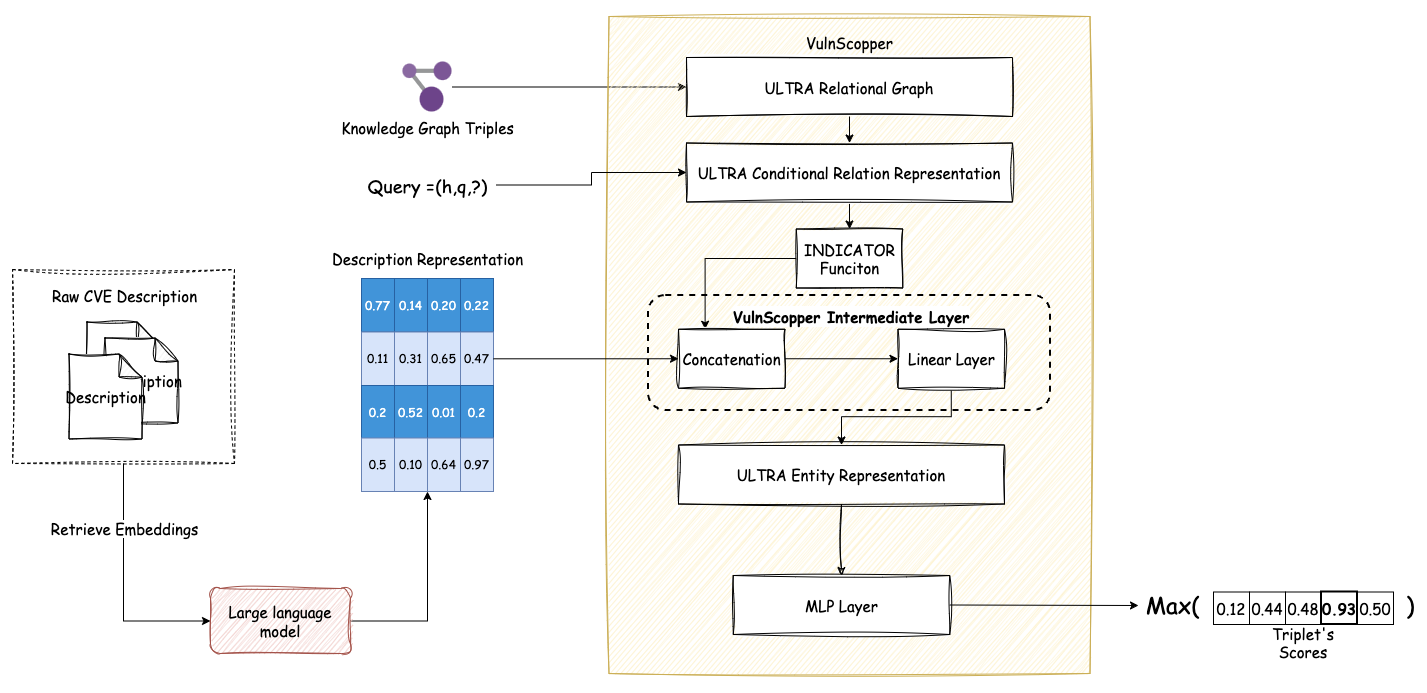}
    \caption{\VulnScopper Multi-modal architecture.}
    \label{fig:model-arch}
\end{figure*}

\VulnScopper consists of four main components to provide link prediction based on the structural information represented by a knowledge graph and the description of security entities. Figure \ref{fig:model-arch} illustrates how these four components are integrated to create VulnScopper. Three of these are ULTRA's components, allowing \VulnScopper to provide link prediction of unseen vulnerabilities, explained in Section \ref{sec:ultra}. 

The fourth component of \VulnScopper is an intermediate component that combines ULTRA's KG representation with the textual description of the entities. \VulnScopper employs a Large Language Model (LLM) to generate the entity description representation and can facilitate the integration of any LLM to create the description-level representation. In this study, we specifically utilize OpenAI's Ada LLM for this purpose. As far as we know, we are the first to integrate a Graph Foundation Model such as ULTRA with an LLM in the security domain and in general. The following sections describe in detail \VulnScopper components and the integration between them.

\subsubsection{VulnScopper Language Model}
In \VulnScopper, language models transform textual descriptions into rich, numerical representations that enhance knowledge graph analytics. Given a graph $G=(E,R,T)$, where $E$ is the set of entities, for each entity $e\in E$, the language model maps the entity's description $d(e)$ to a vector space, resulting in an embedding $h_e\in \mathbb{R}^{1 \times d_t}$ with dimension $d_t$: 
\[
h_e = LLM(d(e))
\]
Specifically, \VulnScopper utilizes OpenAI's Ada\cite{OpenAIAdaEmbeddings} denoted as "text-embedding-ada-002" by OpenAI documentation\cite{OpenAIEmbeddingsModelDocs}. \VulnScopper can flexibly integrate any LLM, as long as $dt\geq d$, where $d$ is ULTRA's feature dimension. This adaptability obtained via a linear layer ensures \VulnScopper's compatibility with various LLMs, as shown in the next section.

\subsubsection{Combining ULTRA with Language Model}
\VulnScopper's core component relies on the textual representation of an entity and its relational representation from ULTRA's second layer, which was introduced in \ref{sec:ultra}. Given a query $q=(h,r,?)$, a relation representation $R_q$ from ULTRA's conditional relation representation step, and a $d_t$-dimensional vector $h_d\in\mathbb{R}^{1xd_t}$ where $d_t \geq d$, of the textual representation of an entity $e\in E$, \VulnScopper obtains a combined node representation of $e$.

\subsection{Implementation and Training}
To implement the components of \VulnScopper, we use Python 3.9 and Pytorch 2.1.0 and integrate with existing ULTRA code. The training process of \VulnScopper follows the standard practice of KG embedding\cite{bordes_transe, distmult, rotate} and adapts the same loss function used by ULTRA, the binary cross entropy loss:
\[
\mathcal{L} = - \log p(u, q, v) - \sum_{i=1}^{n} \frac{1}{n} \log (1 - p(u_i', q, v_i'))
\]

For the relation-level and entity-level GNNs, we adopted the hyperparameters of ULTRA. We conducted a grid search to determine the optimal learning parameters, including learning rate, number of negative examples, epochs, and batches per epoch. Ada's output dimension is 1,536. By concatenating the relation-level GNN output vectors, we obtain a 1,600-dimensional vector. We use two linear layers for the intermediate layer. The first layer has an input size of 1,600 and an output size of 800, while the second layer has an input size of 800 and an output size of 64.

%% file: sections/experiments.tex
In this section, we perform a comprehensive evaluation of \VulnScopper with previous works and other approaches.
We begin by outlining different models used for comparison and their properties, such as whether the model is KG-based, language-based, capable of inductive link prediction, and scalability. Table \ref{table:models_comparison} summarizes the capabilities of different models.

\begin{table*}[h!]
\centering
\begin{tabular}{|l |c | c | r |c | c |}
\hline

\textbf{Model} & \textbf{Paper} & \textbf{KG} & \textbf{LM} & \textbf{Inductive-LP} & \textbf{Scale} \\ \hline \hline
TransE & \cite{CWE-CVE-CPE-Relations} &  \checkmark  &  \texttimes      & \texttimes &  \checkmark \\ \hline
TransE + W2V + GAT & \cite{HybridGat}&  \checkmark  &  \checkmark      & \texttimes  &  \texttimes \\ \hline
Ada + LP      &     $\sim$ \cite{V2W-BERT}  &  \texttimes  &  \checkmark      & \checkmark  &  \checkmark \\ \hline
ChatGPT 4 & \cite{chatgpt_cwe_pred}               &  \texttimes  &  \checkmark      & \checkmark  &  \checkmark \\ \hline
NodePiece  & \cite{galkin2022nodepiece}                 &  \checkmark  &  \texttimes      & \texttimes  &  \checkmark \\ \hline
Inductive-NP & \cite{galkin2022nodepiece}      &  \checkmark  &  \texttimes      & \checkmark  &  \texttimes \\ \hline
ULTRA (All versions)&\cite{galkin2023ultra}      &  \checkmark  &  \texttimes      & \checkmark  &  \texttimes \\ \hline
\hline
NodePiece + LLM &  Our    &  \checkmark  &  \checkmark      & \texttimes  &  \checkmark \\ \hline
\hline
Inductive-NP + LLM & Our & \checkmark  &  \checkmark      & \checkmark  &  \texttimes \\ \hline
\hline
\textbf{\VulnScopper}: ULTRA + LLM    &  Our  &  \checkmark  &  \checkmark      & \checkmark  &  \checkmark \\ \hline

\end{tabular}
\caption{Comparison of the previous works and different models with \VulnScopper.}
\label{table:models_comparison}
\end{table*}

\input{sections/comparisons}

%% file: sections/comparisons.tex
We describe here in detail the different methods used in comparison to our \VulnScopper:

\begin{itemize}
    \item \textbf{TransE} \cite{bordes_transe, CWE-CVE-CPE-Relations}: A knowledge graph embedding method that generates embeddings mapping each entity and relation to a unique vector representation. Each relation between two entities is modeled as a translation from one entity to another utilizing the score function $f(h,r,t)=\lVert h+r-t \rVert$, where $h$  and $t$ are the embeddings of the head and the tail entities, and $r$ is the relation embedding. TransE seeks to minimize this distance for true triples while maximizing it for false ones, differentiating between correct and incorrect knowledge graph triples.
    \item \textbf{TransE + GAT + W2V}\cite{HybridGat}\label{transe_gat_hybrid_explained}: This multi-modal approach combines TransE graph embeddings with word2vec\cite{word2vec}-based description embeddings, employing convolutional neural network to extract features from textual descriptions. The model concat both representations and uses GAT\cite{GAT} to extract the final representations; then, it uses the same score function as utilized by TransE. In \cite{HybridGat}, the author used a graph with 10,502 relations; however, this model struggles to scale with large graphs due to GATs high computational complexity, which grows with the number of relations. This is because GATs calculate attention scores for every pair of connected nodes, leading to a significant increase in computation for graphs with many relations.
    \item \textbf{Ada + LP}: This method uses Ada LLM and a link prediction layer inspired by \cite{V2W-BERT}. First, pairs of entity description embeddings obtained from Ada LLM are sent to the link-prediction component. This component uses a learnable linear output layer with two neurons and softmax activation to classify the combined representations into link and unlink confidence values for CVE-CWE and CPE pairs. The higher link value indicates an association, and the model is optimized by using cross-entropy loss.
    \item \textbf{ChatGPT-4} \cite{chatgpt_cwe_pred}\label{chat_gpt_exp}: This method leverages the ChatGPT-4 LLM chat model. The ChatGPT-4 version used for this experiment is trained with data up to September 2021 and has a context window of 8,192 tokens. Due to its context window limitation, we ask ChatGPT-4 to output only the Top 10 results in a JSON format. We slightly modified the prompts used in \cite{chatgpt_cwe_pred} and obtained better prediction results. Finally, for the CPE prediction, we compare ChatGPT output only to the vendor and product part of the CPE, similarly as shown in Section \ref{sub-sec:vul_graph_construction}. 
    \item \textbf{NodePiece} \cite{galkin2022nodepiece}: This KG embedding method reduces the original large vocabulary of nodes in a graph to a fixed-size, more manageable set by selecting anchor nodes and combining them with relation types to form a NodePiece vocabulary. It tokenizes each node into a hash based on the closest anchors, their distances, and their relational context, which is then encoded into a vector representation using an encoder function.
    \item \textbf{Inductive NodePiece} (Inductive-NP): \cite{galkin2022nodepiece}: This inductive method uses NodePiece's relational context by randomly sampling a set of unique outgoing relations from a node. The relational context is incorporated into the NodePiece vocabulary to construct the embedding vector of each node. A CompGCN\cite{CompGCN} layer is used to improve the representation of nodes in a graph. This layer integrates both node and relation features to enhance graph representation learning. By focusing only on the relational context, an inductive NodePiece can predict links to nodes that have not been seen before.
    \item \textbf{ULTRA} \cite{galkin2023ultra}: ULTRA is described in detail in Section \ref{sec:ultra}. We train two versions of ULTRA from scratch on both graphs in the transductive and inductive mode, denoted as ULTRA NVD, trained on the NVD graph, and ULTRA RH, trained on the Red Hat graph.
    \begin{itemize}
        \item \textbf{ULTRA zero-shot}: This is a pre-trained version of ULTRA, trained on four different graphs. We tested the three provided checkpoints, which include 3, 4, and 50 graphs. Our testing found that using four graphs was the most effective for zero-shot scenarios.
        \item \textbf{ULTRA fine-tuned}: A fine-tuned version of the four graphs checkpoint of ULTRA. We utilized all three checkpoints provided by ULTRA and determined that fine-tuning using the four graph checkpoints achieved the best results. The fine-tuned ULTRA on the NVD graph is denoted as ULTRA NVD tuned, and for the Red Hat graph, ULTRA RH tuned.
    \end{itemize}
\end{itemize}

In the search for an integrated representation of the entity graph and textual description, we introduce two multi-modal models: NodePiece+LLM and Inductive-NP + LLM:

\begin{itemize}
    \item \textbf{NodePiece + LLM}: A modified version of NodePiece that utilizes the intermediate layer used by \VulnScopper with the entity embeddings obtained from NodePiece and the description embeddings obtained from Ada LLM.
    \item \textbf{Inductive NodePiece + LLM} (Inductive-NP + LLM)\cite{galkin2022nodepiece}: A modified version of the inductive-NodePiece that utilizes the intermediate layer used by \VulnScopper with the entity embeddings obtained from NodePiece and the description embeddings obtained from Ada LLM. The entity embedding obtained from NodePiece's relational context is used in this model, distinguishing it from NodePiece + LLM.
\end{itemize}

%% file: sections/evaluation.tex
\subsection {Evaluation Metrics}
We evaluate all the models described in the previous section and use the standard KG link prediction metrics to report their accuracy. 

The MRR is a statistical measure for evaluating the quality of a set of ranked results. It is the average of the reciprocal ranks of results for a set of queries. The reciprocal rank of a query response is the multiplicative inverse of the rank of the first correct answer:

\[
MRR = \frac{1}{|Q|} \sum_{i=1}^{|Q|} \frac{1}{rank_i}
\]
where $|Q|$ is the number of queries, and $rank_i$ is the rank of the first correct answer for the $i$-th query. 

The MRR of ChatGPT is marked with an "X" because its token limitation necessitates restricting it to rank only the top 10 entities, rendering its MRR incomparable to other models that rank all entities.\label{chat-gpt-mrr}

Hits@K measures the proportion of correct predictions that appear in the top $K$ ranked items. It is defined as:

\[
Hits@K = \frac{1}{|Q|} \sum_{q=1}^{|Q|} I(rank_q \leq K)
\]
where $|Q|$ is the total number of queries, ${I}(\cdot)$ is the indicator function that is 1 if the correct answer is within the top $K$ positions for the $q$-th query and 0 otherwise, and $rank_q$ is the rank position of the correct answer for the $q$-th query.

\subsection{Datasets}
\input{sections/datasets}

%% file: sections/datasets.tex
We construct datasets for our experiments based on the methodology outlined in Section \ref{sub-sec:vul_graph_construction}. In the transductive setup, triples for validation and test graphs are randomly selected, ensuring all entities are also presented in the training graph. Both graphs utilize all the available data in each dataset.
\begin{table}[h!]
\centering
\begin{tabular}{lccc}
\hline
\textbf{Dataset} & \textbf{\#Train} & \textbf{\#Validation} & \textbf{\#Test} \\ \hline
NVD      & 1,037,927 &       3,021      & 11,475   \\ \hline
RedHat   &  107,901  &       501      & 1379   \\ \hline
\end{tabular}
\caption{The number of triples in the train, validation, and test graphs for each dataset in the transductive setup.}
\label{table:graph_entities_count}
\end{table}

In the inductive setup, we divide the graph temporally and include all data available in each dataset up to October 2023. The training graph contains triples up to January 1st, 2023. Subsequent relations created after January 1st, 2023, and up to October 18th, 2023, are allocated to the inference, validation, and test graphs.
\begin{table}[h!]
\centering
\begin{tabular}{lcrcc}
\hline
\textbf{Dataset} & \textbf{\#Train} & \textbf{\#Inference} & \textbf{\#Validation} & \textbf{\#Test} \\ \hline
NVD     &   880,758 & 166,799       &       1,461      & 3,405   \\ \hline
RedHat  &   88,531 & 20,258       &      300       & 692   \\ \hline
\end{tabular}
\caption{The number of triples in the train, validation, and test graphs for each dataset in the inductive setup.}
\label{table:graph_entities_count}
\end{table}

%% file: sections/results.tex
\subsection{Transductive Link Prediction Results}
\label{sec:transductive_lp_res}
Tables \ref{table:transductive_redhat_cve_cwe_res} and \ref{transductive_redhat_cve_cpe_res} present the CVE to CWE and CVE to CPE transductive link prediction results on the Red Hat dataset. The complete results are in the Appendix \ref{apendix:transductive_lp_res}. The evaluation results indicate that \VulnScopper has performed better than all the other methods in predicting the link between CPE and CVE on both datasets, with 76\% and 71\% Hits@10 on Red Hat and NVD datasets, respectively. 

In the CVE to CWE link prediction task, \VulnScopper outperformed all other methods when evaluated on the RedHat dataset with 70\% Hits@10. In the NVD dataset, \VulnScopper has achieved the best MRR and Hits@10 of 71\%, while ChatGPT has performed better on Hits@3 and Hits@1 with 59\% and 49\%, respectively.

\begin{table}[h!]
\centering
\begin{tabular}{lcccc}
\hline
\multirow{2}{*}{Model} &   \multirow{2}{*}{MRR} & \multicolumn{3}{c}{Hit@k} \\ \cline{3-5} 
                       &                      & @10           & @3            & @1            \\ \hline
TransE                 & 0.204                & 0.431         & 0.219         & 0.099         \\ 
TransE + Gat + W2V     & 0.311                & 0.531         & 0.222         & 0.145         \\
ADA + LP               & 0.114                & 0.355         & 0.201         & 0.133         \\
ChatGPT 4            & X                    & 0.518         & 0.427          & 0.344         \\
Nodepiece              & 0.173                & 0.402         & 0.178         & 0.070         \\
Nodepiece + LLM        & 0.325                & 0.551         & 0.420         & 0.248         \\
Inductive NP    & 0.200                & 0.435         & 0.276         & 0.124         \\
Inductive NP + LLM & 0.349             & 0.650         & 0.443         & 0.250         \\
ULTRA Zero-Shot        & 0.270                 & 0.634         & 0.290          & 0.128         \\
ULTRA (tuned)   & 0.372                & 0.696         & 0.448         & 0.360         \\
\textbf{\VulnScopper}         & \textbf{0.382}                & \textbf{0.701}         & \textbf{0.469}         & \textbf{0.377}         \\
ULTRA             & 0.344                & 0.586         & 0.324         & 0.247         \\
\end{tabular}
\caption{Transductive link-prediction of CVE to CWE on RedHat Dataset.  ChatGPT 4 MRR is marked with "X" since it returns
only the top 10 results.}
\label{table:transductive_redhat_cve_cwe_res}
\end{table}

\begin{table}[h!]
\centering
\begin{tabular}{lcccc}
\hline
\multirow{2}{*}{Model} & \multirow{2}{*}{MRR} & \multicolumn{3}{c}{Hit@k} \\ \cline{3-5} 
                       &                      & @10           & @3            & @1            \\ \hline
TransE                 & 0.373                & 0.556         & 0.416         & 0.277         \\
TransE + Gat + W2V     & 0.319                & 0.488         & 0.364         & 0.210         \\
ADA + LP               & 0.201                & 0.188         & 0.167         & 0.125         \\
ChatGPT 4            & X                & 0.153         & 0.141         & 0.116         \\
Nodepiece              & 0.178                & 0.324         & 0.196         & 0.102         \\
Nodepiece + LLM        & 0.203                & 0.411         & 0.213         & 0.111         \\
Inductive NP    & 0.366                & 0.514         & 0.332         & 0.269         \\
Inductive NP + LLM & 0.417             & 0.552         & 0.437         & 0.348         \\
ULTRA Zero-Shot        & 0.198                & 0.416         & 0.197         & 0.105         \\
ULTRA (tuned)   & 0.581                & 0.756         & 0.648         & 0.480          \\
\textbf{\VulnScopper}         & \textbf{0.622}                & \textbf{0.768}         & \textbf{0.652}         & \textbf{0.533}         \\
ULTRA               & 0.522                & 0.701         & 0.634         & 0.460         \\
\end{tabular}
\caption{Transductive link-prediction of CVE to CPE on
RedHat Dataset. ChatGPT 4 MRR is marked with "X" since
it returns only the top 10 results}
\label{transductive_redhat_cve_cpe_res}
\end{table}

\subsection{Inductive Link Prediction}
\label{sec:inductive_lp_res}
Next, for the inductive link prediction task, we evaluate 8 inductive model configurations from the inductive models summarized in Table \ref{table:models_comparison}. Tables \ref{table:inductive_nvd_cve_cpe_res} and \ref{inductive_nvd_cve_cwe_res} present the inductive link prediction results for the NVD Dataset on the CPE and CWE prediction tasks. The complete inductive link prediction results are shown in Appendix \ref{apendix:inductive_lp_res}. 

Specifically, \VulnScopper performs better in CPE prediction on the NVD dataset, achieving Hits@10 of 60\%.
ULTRA fine-tuned for the Red Hat dataset outperforms the other methods with Hits@10 of 66\% and 78\% on CWE and CPE predictions, respectively on the Red Hat dataset. Finally, for CWE prediction on the NVD dataset, ChatGPT 4 gets the best results with 74\% Hits@10.

\begin{table}[h!]
\centering
\begin{tabular}{lcccc}
\hline
 \multirow{2}{*}{Model} &  \multirow{2}{*}{MRR} & \multicolumn{3}{c}{Hit@k} \\ 

\cline{3-5} 
                       &                      & @10           & @3            & @1  
                       \\ \hline
ADA + LP               & 0.205                & 0.233         & 0.180          & 0.154         \\
ChatGPT 4            & X                    & 0.162         & 0.139         & 0.112         \\
Inductive NP   & 0.275                & 0.469         & 0.288         & 0.203         \\
Inductive NP + LM & 0.318              & 0.556         & 0.340         & 0.220         \\
ULTRA Zero-Shot        & 0.320                 & 0.392         & 0.354         & 0.269         \\
ULTRA (tuned)  & 0.535                & 0.569         & 0.528         & 0.460          \\
\textbf{\VulnScopper}        & \textbf{0.547}                & \textbf{0.603}         & \textbf{0.540}         & \textbf{0.486}         \\
ULTRA              & 0.401                & 0.466         & 0.320          & 0.226         \\ \hline
\end{tabular}
\caption{Inductive link-prediction of CVE to CPE on NVD.  ChatGPT 4 MRR is marked with "X" since it returns
only the top 10 results.}
\label{table:inductive_nvd_cve_cpe_res}
\end{table}

\begin{table}[h!]
\centering
\begin{tabular}{lcccc}
\hline
\multirow{2}{*}{Model} & \multirow{2}{*}{MRR} & \multicolumn{3}{c}{Hit@k} \\ \cline{3-5} 
                       &                      & @10           & @3            & @1            \\ \hline
ADA + LP               & 0.209                & 0.433         & 0.301         & 0.206         \\
\textbf{ChatGPT 4}            & X                    & \textbf{0.744} & \textbf{0.637} & \textbf{0.529} \\
Inductive Nodepiece    & 0.191                & 0.441         & 0.253         & 0.123         \\
Inductive Nodepiece + LM & 0.214              & 0.458         & 0.237         & 0.155         \\
ULTRA Zero-Shot        & 0.247                & 0.452         & 0.350         & 0.089         \\
ULTRA (Finetuned)  & 0.430                & 0.639         & 0.566         & 0.311         \\
VulnScopper        & 0.452                & 0.675         & 0.534         & 0.472         \\
ULTRA              & 0.342                & 0.498         & 0.270         & 0.185         \\ \hline
\end{tabular}
\caption{Inductive link-prediction of CVE to CWE on the NVD Dataset. ChatGPT 4 MRR is marked with "X" since it returns only the
top 10 results.}
\label{inductive_nvd_cve_cwe_res}
\end{table}

\subsection{ CWE Categorization  per Dataset}
In our examination of CWE categorization performance, it becomes evident that ChatGPT achieved better results on the NVD dataset compared to the Red Hat CVE dataset, as shown in Tables \ref{inductive_nvd_cve_cwe_res}, \ref{inductive_redhat_cve_cwe_res}, \ref{transductive_nvd_cve_cwe_res}, \ref{table:transductive_redhat_cve_cwe_res}. This disparity is further analyzed in Section \ref{sub-sec:cve_cwe_association_analysis}, where we delve into the distinct CWE categorization used by both databases.

Table \ref{table:chat_gpt_bias} presents the evaluation results of ChatGPT 4 on the Red Hat CVE dataset. The column CWE specifies the categorization scheme used, either NVD or Red Hat, and the mode column determines whether the inductive dataset (I) or the transductive dataset (T) was used in the evaluation. The results show that ChatGPT 4 prefers the NVD's classification scheme. This outcome emphasizes the necessity of a specialized tool like VulnScopper, which benefits from training on targeted datasets to mitigate such biases.

\begin{table}[h!]
\centering
\begin{tabular}{lllccc}
\hline
\multirow{2}{*}{Model} & \multirow{2}{*}{Mode} & \multirow{2}{*}{ Label} & \multicolumn{3}{c}{Hit@k} \\ \cline{4-6} 
             &          &                      & @10           & @3            & @1            \\ \hline
ChatGPT 4 & T & Red Hat  & 0.518  & 0.427     & 0.344         \\
\VulnScopper & T & Red Hat     & 0.701  & 0.469     & 0.377         \\ \hline
ChatGPT 4 & T & NVD     & 0.643   & 0.506    & 0.360         \\ \hline
ChatGPT 4  & I &Red Hat      & 0.543  & 0.370     & 0.308         \\
\VulnScopper & I & Red Hat     & 0.648  & 0.376     & 0.351         \\
\hline
ChatGPT 4 & I & NVD          & 0.583  & 0.451    & 0.334       \\  \hline

\end{tabular}
\caption{Evaluation of ChatGPT 4 for CWE categorization on Red Hat CVE Dataset, highlighting the contrast between NVD and Red Hat classification styles and emphasizing ChatGPT's preference towards NVD categorization.}

\label{table:chat_gpt_bias}
\end{table}

%% file: sections/examples.tex
We illustrate the effectiveness of \VulnScopper by examining recent vulnerability cases. This section covers two well-known vulnerability cases: CVE-2023-4863 in 'libwebp' and CVE-2023-5217 in 'libvpx'. We used the \VulnScopper model trained on NVD data up to 2023 to predict the vulnerable CPE configurations of the three vulnerability cases. 

We leverage each vulnerability case's description as input to \VulnScopper, including all its known CPE configurations in the inference graph. The aim is to identify new links to CPE configurations not already in the graph, thus uncovering potentially vulnerable ones. Based on \VulnScopper's ranked output of CPE configurations, we remove CPEs previously identified in the NVD (true links) and subject the top 10 outcomes to thorough manual review and cross-validation against additional security databases. This process highlights \VulnScopper's capability to streamline analysts' workloads by focusing on specific vulnerable products, consequently aiming to decrease the risk period for software systems.

\subsection{CVE-2023–4863 - 'libwebp'}
CVE-2023–4863 is a heap buffer overflow vulnerability in open-source libwebp. Many web browsers and image editors use this library to display WebP format images. Initially, this CVE in NVD was exclusively linked to Google Chrome. However, its scope expanded as other browsers began reporting similar concerns. Based on the information available in NVD for this vulnerability, the following CPE configurations are vulnerable:

\begin{itemize}
    \item \texttt{cpe:2.3:a:google:chrome}
    \item \texttt{cpe:2.3:o:fedoraproject:fedora}
    \item \texttt{cpe:2.3:o:debian:debian\_linux}
    \item \texttt{cpe:2.3:a:mozilla:firefox}
    \item \texttt{cpe:2.3:a:mozilla:firefox\_esr}
    \item \texttt{cpe:2.3:a:mozilla:thunderbird}
    \item \texttt{cpe:2.3:a:microsoft:edge}
    \label{list:exmple1}
\end{itemize}

Table \ref{table:cve-2023-4863_example} illustrates VulnScopper's ability to identify previously unrecognized CPE configurations. Within the top 10 predictions, we find 8 CPE configurations that could be validated through cross-checking.

\begin{table}[h!]
\centering
\begin{tabular}{lcr}
\hline
\textbf{CPE Configuration} & \textbf{Rank} & \textbf{Ref} \\ \hline
cpe:2.3:a:microsoft:edge\_chromium  &     1   &       \cite{CVE-2023-4863_edge_chromium}   \\ \hline
cpe:2.3:a:imagemagick:imagemagick   &     2   &       \cite{CVE-2023-4863_snyk}   \\ \hline
cpe:2.3:o:canonical:ubuntu\_linux   &     4   &       \cite{CVE-2023-4863_ubuntu}   \\ \hline
cpe:2.3:a:ffmpeg:ffmpeg             &     5   &       \cite{CVE-2023-4863_snyk}   \\ \hline
cpe:2.3:o:redhat:enterprise\_linux  &     6   &       \cite{CVE-2023-4863_redhat}   \\ \hline
cpe:2.3:a:libtiff:libtiff           &     7   &           \cite{CVE-2023-4863_rezilion}   \\ \hline
cpe:2.3:a:qt:qt                     &     8   &       \cite{CVE-2023-4863_qt}   \\ \hline
cpe:2.3:a:python:pillow             &     10   &      \cite{CVE-2023-4863_snyk}   \\ \hline
\end{tabular}
\caption{Vulnerable CPE configuration of CVE-2023-4863 among the top 30 \VulnScopper predicitons.}
\label{table:cve-2023-4863_example}
\end{table}

\subsection{CVE-2023-5217 - 'libvpx'}
\label{sub:case-study-libvpx}
Subsequently, we examine CVE-2023-5217, a high-impact heap buffer overflow vulnerability mainly affecting web browsers utilizing libvpx software library. The vulnerable CPE configurations for this vulnerability are shown in Appendix \ref{appendix:libvpx_vulnerable_cpes}.

Table \ref{table:cve-2023-5217_example} illustrates VulnScopper's ability to identify previously unrecognized CPE configurations. Within the top 10 predictions, 7 CPE configurations were validated through cross-checking.

\begin{table}[h!]
\centering
\begin{tabular}{lcr}
\hline
\textbf{CPE Configuration} & \textbf{Rank} & \textbf{Ref} \\ \hline
cpe:2.3:o:redhat:enterprise\_linux   &     1   &       \cite{CVE-2023-4863_redhat_Advisory}   \\ \hline
cpe:2.3:a:qt:qt                     &     3   &       \cite{CVE-2023-5217_qt}   \\ \hline
cpe:2.3:a:torproject:tor            &     4   &       \cite{CVE-2023-5217_tor_browser}   \\ \hline
cpe:2.3:a:brave:brave               &     5   &       \cite{CVE-2023-5217_brave1, CVE-2023-5217_brave2}   \\ \hline
cpe:2.3:a:ffmpeg:ffmpeg             &     6   &       \cite{CVE-2023-5217_ffmpeg_rezilion}   \\ \hline
cpe:2.3:a:electronjs:electron       &     8   &           \cite{CVE-2023-5217_electron}   \\ \hline
cpe:2.3:a:cefsharp:cefsharp         &     9   &      \cite{CVE-2023-5217_cefsharp}   \\ \hline
\end{tabular}
\caption{Vulnerable CPE configuration of CVE-2023-4863 among the top 30 \VulnScopper predicitons.}
\label{table:cve-2023-5217_example}
\end{table}

%% file: sections/discussion.tex
\emph{Address unseen vulnerabilities}. The results and case studies demonstrate ULTRA's ability to manage unseen vulnerabilities, eliminating the need for constant model retraining by a domain expert for each new vulnerability report.

\emph{Scalability}. Utilizing ULTRA, VulnScopper enables the prediction of CWEs and CPEs for unseen vulnerabilities across extensive graphs. Section 3 discusses the deployment configuration that allows VulnScopper to make predictions based on all the CVE, CWE, and CPE entities in NVD and Red Hat databases up to 2023.

\emph{Performance and Accuracy}. Our exploration has underscored the importance of a multi-modal KG and NLP approach in comprehending the scope of vulnerabilities. Language models, which treat each CVE as an isolated instance, often overlook the intricate CPE configurations, mainly when such details are not explicit in the descriptions. Although adept at providing structural insight, KG models cannot grasp the contextual nuances within textual descriptions, a shortfall that becomes particularly conspicuous in predicting CWEs where such context is crucial. VulnScopper's multi-modal methodology, which integrates the structural knowledge of KGs with the descriptive understanding of language models, addresses this gap.

\emph{AI-Assisted System}. The ability of VulnScopper to rank links to unseen vulnerabilities is a step forward in an AI-assisted system that helps domain experts analyze vulnerabilities, prioritize, and respond to threats, significantly accelerating the vulnerability scoping process. 

%% file: sections/conclusions.tex
In this paper, we present \VulnScopper, a robust system that harnesses the synergy of Graph Foundation Models and Large Language Models to revolutionize the prediction and analysis of vulnerabilities. By adeptly combining the structural insights from vulnerability databases with the nuanced contextual understanding of language models, \VulnScopper achieves exceptional accuracy in transductive and inductive settings, discerning intricate patterns and linkages between CVEs, CWEs, and CPEs. Therefore, \VulnScopper is a step towards an AI-assisted system that can help analysts analyze vulnerabilities efficiently.

\VulnScopper not only elevates the precision of identifying relationships in cyber-physical systems but also illuminates the path for constructing a general-purpose cybersecurity multi-modal graph and text-based foundation model. This promising direction for future research has the potential to create a unified framework for cyber threat intelligence, further empowering the detection, prevention, and mitigation strategies in the ever-evolving landscape of cybersecurity threats.

\section*{Acknowledgment}	
This research was partly supported by RedHat Research. 

%% file: sections/appendix.tex
\subsection{Transductive link-prediction results}
\label{apendix:transductive_lp_res}
In section \ref{sec:transductive_lp_res} Tables \ref{table:transductive_redhat_cve_cwe_res} and \ref{transductive_redhat_cve_cpe_res}, we show the results of the CVE to CWE, and CVE to CPE link prediction for the Red Hat dataset. For the NVD dataset, Tables \ref{transductive_nvd_cve_cwe_res} and \ref{transductive_nvd_cve_cpe_res} show the results for CVE to CWE and CPE, respectively.

\begin{table}[h!]
\centering
\begin{tabular}{lcccc}
\hline
\multirow{2}{*}{Model} & \multirow{2}{*}{MRR} & \multicolumn{3}{c}{Hit@k} \\ \cline{3-5} 
                       &                      & @10           & @3            & @1            \\ \hline
TransE                 & 0.206                & 0.454         & 0.227         & 0.102         \\
TransE + Gat + W2V     & -                    & -             & -             & -             \\
ADA + LP               & 0.188                & 0.431         & 0.255         & 0.137         \\
ChatGPT 4            & X                    & 0.699         & \textbf{0.597}         & \textbf{0.494}         \\
Nodepiece              & 0.213                & 0.360          & 0.219         & 0.109         \\
Nodepiece + LLM        & 0.321                & 0.570         & 0.412         & 0.255         \\
Inductive NP    & 0.180                & 0.305         & 0.244         & 0.111         \\
Inductive NP + LLM & 0.404             & 0.608         & 0.435         & 0.283         \\
ULTRA Zero-Shot        & 0.205                & 0.424         & 0.263         & 0.065         \\
ULTRA (tuned)  & 0.372                & 0.643         & 0.422         & 0.241         \\
\textbf{\VulnScopper}        & \textbf{0.424}                & \textbf{0.714}         & 0.483         & 0.285         \\
ULTRA              & 0.293                & 0.498         & 0.270         & 0.165         \\ \hline
\end{tabular}
\caption{Transductive link-prediction of CVE to CWE on NVD Dataset.  Prediction results of TransE + Gat + W2V are denoted by a "-" to reflect that a model did not scale to encompass the entire dataset. ChatGPT 4 MRR is marked with "X" since it returns only the
top 10 results. }
\label{transductive_nvd_cve_cwe_res}
\end{table}

\begin{table}[h!]
\centering
\begin{tabular}{lcccc}
\hline
\multirow{2}{*}{Model} & \multirow{2}{*}{MRR} & \multicolumn{3}{c}{Hit@k} \\ \cline{3-5} 
                       &                      & @10           & @3            & @1            \\ \hline
TransE                 & 0.454                & 0.595         & 0.489         & 0.383         \\
TransE + Gat + W2V     & -                    & -             & -             & -             \\
ADA + LP               & 0.114                & 0.207         & 0.183         & 0.153         \\
ChatGPT 4            & X                    & 0.180         & 0.157         & 0.127         \\
Nodepiece              & 0.385                & 0.501          & 0.335         & 0.201         \\
Nodepiece + LLM        & 0.399                & 0.590         & 0.353         & 0.244         \\
Inductive NP    & 0.378                & 0.524         & 0.406         & 0.320         \\
Inductive NP + LLM & 0.442             & 0.552         & 0.437         & 0.348         \\
ULTRA Zero-Shot        & 0.220                & 0.275         & 0.232         & 0.179         \\
ULTRA (tuned)  & 0.523                & 0.643         & 0.540         & 0.462         \\
\textbf{\VulnScopper}        & \textbf{0.590}                & \textbf{0.715}         & \textbf{0.638}         & \textbf{0.526}         \\
ULTRA              & 0.488                & 0.599         & 0.340         & 0.282         \\ \hline
\end{tabular}
\caption{Transductive link-prediction of CVE to CPE on NVD Dataset.  Prediction results of TransE + Gat + W2V are denoted by a "-" to reflect that a model did not scale to encompass the entire dataset. ChatGPT 4 MRR is marked with "X" since it returns only the top 10 results for each configuration. }
\label{transductive_nvd_cve_cpe_res}
\end{table}

\subsection{Inductive link-prediction results}
\label{apendix:inductive_lp_res}
In section \ref{sec:inductive_lp_res} Tables \ref{table:inductive_nvd_cve_cpe_res} and \ref{inductive_nvd_cve_cwe_res}, we show the results of the CVE to CPE and CVE to CWE link prediction results for the NVD dataset. For the Red Hat dataset, Tables \ref{inductive_redhat_cve_cwe_res} and \ref{inductive_redhat_cve_cpe_res} show the results for CVE to CWE and CPE, respectively.

\begin{table}[h!]
\centering
\begin{tabular}{lcccc}
\hline
\multirow{2}{*}{Model} & \multirow{2}{*}{MRR} & \multicolumn{3}{c}{Hit@k} \\ \cline{3-5} 
                       &                      & @10           & @3            & @1            \\ \hline
ADA + LP               & 0.114                & 0.401         & 0.261         & 0.173         \\
ChatGPT 4            & X                    & 0.543         & 0.370         & 0.308         \\
Inductive Nodepiece    & 0.188                & 0.411         & 0.253         & 0.104         \\
Inductive Nodepiece + LM & 0.247              & 0.435         & 0.329         & 0.122         \\
ULTRA Zero-Shot        & 0.198                & 0.530         & 0.172         & 0.080         \\
\textbf{ULTRA (tuned)}   & \textbf{0.363}       & \textbf{0.660} & \textbf{0.413} & \textbf{0.363} \\
VulnScopper         & 0.348                & 0.648         & 0.376         & 0.351         \\
ULTRA               & 0.330                & 0.617         & 0.339         & 0.191         \\
\end{tabular}
\caption{Inductive link-prediction of CVE to CWE on the Red Hat Dataset. ChatGPT 4 MRR is marked with "X" since it returns only the
top 10 results.}
\label{inductive_redhat_cve_cwe_res}
\end{table}

\begin{table}[h!]
\centering
\begin{tabular}{lcccc}
\hline
\multirow{2}{*}{Model} & \multirow{2}{*}{MRR} & \multicolumn{3}{c}{Hit@k} \\ \cline{3-5} 
                       &                      & @10           & @3            & @1            \\ \hline
ADA + LP               & 0.189                & 0.204         & 0.133         & 0.101         \\
ChatGPT 4            & X                    & 0.333         & 0.298         & 0.229         \\
Inductive Nodepiece    & 0.280                & 0.474         & 0.302         & 0.214         \\
Inductive Nodepiece + LM & 0.318              & 0.581         & 0.369         & 0.215         \\
ULTRA Zero-Shot        & 0.309                & 0.469         & 0.309         & 0.232         \\
\textbf{ULTRA (tuned)}   & \textbf{0.611}       & \textbf{0.780} & \textbf{0.669} & \textbf{0.519} \\
VulnScopper        & 0.587                & 0.761         & 0.652         & 0.491         \\
ULTRA            & 0.564                & 0.741         & 0.634         & 0.460         \\
\end{tabular}
\caption{Inductive link-prediction of CVE to CPE on the Red Hat Dataset. ChatGPT 4 MRR is marked with "X" since it returns only the
top 10 results.}
\label{inductive_redhat_cve_cpe_res}
\end{table}

\subsection{Vulnerable CPE configurations}
\subsubsection{CVE-2023-5217 'libvpx'}
\label{appendix:libvpx_vulnerable_cpes}
We analyze CVE-2023-5217 'libvpx' in Section \ref{sub:case-study-libvpx}. Based on the information available in NVD for this vulnerability, the following CPE configurations are vulnerable:

\begin{itemize}
    \item \texttt{cpe:2.3:a:webmproject:libvpx}
    \item \texttt{cpe:2.3:a:google:chrome}
    \item \texttt{cpe:2.3:a:mozilla:firefox}
    \item \texttt{cpe:2.3:a:mozilla:firefox\_esr}
    \item \texttt{cpe:2.3:a:mozilla:firefox\_focus}
    \item \texttt{cpe:2.3:a:mozilla:thunderbird}
    \item \texttt{cpe:2.3:a:microsoft:edge}
    \item \texttt{cpe:2.3:o:fedoraproject:fedora}
    \item \texttt{cpe:2.3:o:debian:debian\_linux}
    \item \texttt{cpe:2.3:o:apple:ipad\_os}
    \item \texttt{cpe:2.3:o:apple:iphone\_os}
    \label{list:exmple2}
\end{itemize}